# On the mechanisms of superfluidity in atomic nuclei

A. V. Avdeenkov and S. P. Kamerdzhiev*)

*Physics and Power Engineering Institute State Science Center of the Russian Federation, 249020 Obninsk, Kaluga Region, Russia*

A system of equations is obtained for the Cooper gap in nuclei. The system takes two mechanisms of superfluidity into account in an approximation quadratic in the phonon-production amplitude: a Bardeen–Cooper–Schrieffer (BCS)-type mechanism and a quasiparticle–phonon mechanism. These equations are solved for $^{120}$Sn in a realistic approximation. If the simple procedures proposed are used to determine the new particle–particle interaction and to estimate the average effect, then the contribution of the quasiparticle–phonon mechanism to the observed width of the pairing gap is 26% and the BCS-type contribution is 74%. This means that at least in semimagic nuclei pairing is of a mixed nature — it is due to the two indicated mechanisms, the first being mainly a surface mechanism and the second mainly a volume mechanism. © *1999 American Institute of Physics.*
[S0021-3640(99)00110-3]

In the microscopic theory of ordinary superconductors the Éliashberg theory,[1] in which the interaction of the electrons that leads to pairing is due to the exchange of phonons, describes the mechanism of superconductivity quite well. In the weak electron–phonon interaction limit $g^2 \ll 1$ this mechanism reduces to the well-known Bardeen–Cooper–Schrieffer (BCS) model.

The situation is somewhat different in the microscopic theory of nuclei with pairing (nonmagic nuclei). As a rule, the width of the superfluid gap is determined experimentally or calculated using the BCS equation with a phenomenologically chosen particle–particle ($pp$) interaction.[2,3] This interaction and therefore the gap are energy-independent. In other words the quasiparticle–phonon interaction (QPI) in the problem of pairing in nuclei is taken into account only effectively — to the extent that the quasiparticle–phonon pairing mechanism can be reduced to the indicated BCS mechanism. This would be justified if $g^2 \ll 1$, where $g$ is the phonon production amplitude, in nuclei. In nuclei with pairing, however, $g^2 > 1$ can occur in each of the two nucleon systems because of the existence of a low-lying $2^+$ collective level.[4] In nuclei with pairing the weak inequality $g^2 < 1$ can occur in one of the nucleon systems (semimagic nuclei; see calculations in Ref. 5 for $^{120}$Sn). Therefore the quasiparticle–phonon pairing mechanism must be examined explicitly, and it is of interest to study the realistic case $g^2 < 1$ first.

It is well known that the most highly collective low-lying phonons, which make the largest contribution to QPI effects in nuclei, are mainly surface oscillations. For this reason, singling out the quasiparticle–phonon pairing mechanism explicitly will make it possible to answer the old question of whether pairing in nuclei is a volume or surface effect. This question has been discussed in Ref. 6 at the phenomenological level — introducing the ''internal'' $F_{in}^{\xi}$ and ''external'' $F_{ex}^{\xi}$ pp-interaction amplitudes — on the basis of the theory of finite Fermi systems, where it was found that for Sn isotopes pairing is primarily a volume effect. The question of the nature of pairing has been raised in Ref. 7 at the microscopic level.

It has been found that it is important to take into account the QPI in the particle–hole channel in order to gain a quantitative and a qualitative understanding of many nuclear phenomena, above all for describing excitations of nuclei.[3,4,8] A systematic allowance for the QPI in the pp channel, including for the pairing problem, should improve the description of at least the low-lying excitations in odd-mass nuclei (see Ref. 5, where this is shown quantitatively) and in even–even nuclei with pairing. This is especially important now in connection with the advent of qualitatively new experimental possibilities in EUROBALL $\gamma$ spectrometers, which are now in operation in Europe and the USA.[9]

In Fermi systems with superfluidity it is necessary to use, besides the standard single-particle Green's functions $G$ and $G^{(h)}$, the anomalous single-particle Green's functions $F^{(1)}$ and $F^{(2)}$. For a realistic description the well-known components, i.e., the mean field and pairing, described by a BCS-type equation, should be singled out explicitly, after which corrections to first order in $g^2$ in all mass operators should be studied. This problem, i.e., the formulation of the gap equations in the $g^2$ approximation, is studied in the first part of this letter. The first computational results for $^{120}$Sn are presented in the second part.

We shall represent each of the complete mass operators in the system of equations for $G$ and $F$ as a sum of two terms, the first being energy-independent and the second energy-dependent

$$\Sigma(\varepsilon) = \tilde{\Sigma} + M(\varepsilon), \quad \Sigma^{(h)}(\varepsilon) = \tilde{\Sigma}^{(h)} + M^{(h)}(\varepsilon), \tag{1}$$

$$\Sigma^{(1)}(\varepsilon) = \tilde{\Sigma}^{(1)} + M^{(1)}(\varepsilon), \quad \Sigma^{(2)}(\varepsilon) = \tilde{\Sigma}^{(2)} + M^{(2)}(\varepsilon),$$

where $\tilde{\Sigma}$ and $\tilde{\Sigma}^{(h)}$ correspond to the mean field and $\tilde{\Sigma}^{(1)}$ and $\tilde{\Sigma}^{(2)}$ correspond to pairing, described by a BCS-type mechanism. The quantities $M^i$ contain the QPI explicitly and are taken in the $g^2$ approximation:

$$M(\varepsilon) = M^{(h)}(-\varepsilon) = \text{[diagram]}, \tag{2}$$

$$M^{(1)}(\varepsilon) = \text{[diagram]}, \quad M^{(2)}(\varepsilon) = \text{[diagram]}, \tag{3}$$

where a circle denotes the phonon production amplitude $g$ and the Green's functions in the mass operators $M^i$ do not contain phonons. Pair phonons are neglected here and below, since their contribution should be small.

In what follows it should be kept in mind that the initial components of the problem are the mean field, described by the phenomenological Woods–Saxon potential, and the gap width, which satisfies the BCS equation with a phenomenologically chosen *pp* interaction. At present this approach is the most realistic in the theory of nuclei, especially for describing experiments for nonmagic nuclei.

On account of the phenomenological nature of the input quantities — the single-particle energies $\varepsilon_\lambda$ and the gap widths $\Delta_\lambda$ ($\lambda$ denotes the single-particle quantum numbers), the energies $\varepsilon_\lambda$ should contain a contribution from the terms $M$,[10] and the quantities $\Delta_\lambda^{(1),(2)}$ should contain a contribution due to the terms $M^{(1),(2)}$. The latter can be seen from the standard BCS equation with a phenomenological *pp* interaction, written in terms of the Green's functions method,[2]

$$\Delta_\lambda^{(2)} = \Delta_\lambda^{(1)} = \sum_{\lambda'} F^\xi_{\lambda \bar\lambda \lambda' \bar{\lambda'}} F^{(2)}_{\lambda'}; \tag{4}$$

where $F^{(2)}_{\lambda'}$ is the anomalous Gor'kov–Green's function and $F^\xi$ is the renormalized interaction amplitude, which is irreducible in the *pp* channel. Therefore $F^\xi$ contains diagrams corresponding to phonon exchange, i.e., we can write[7,11] (symbolically)

$$F^\xi = W + gDg, \tag{5}$$

where $W$ is a new *pp* interaction and $D$ is the phonon Green's function. Then, singling out the pole diagram with a phonon in Eq. (5), according to Eq. (4), corresponds to taking $M^{(2)}$ from Eq. (3) into account in Eq. (1). Therefore, in order to avoid taking the quantities $M^i$ into account twice they must be excluded from the phenomenological quantities, i.e., the QPI must be ''removed'' from the latter. These refined quantities are everywhere marked with a tilde.

The system of equations for the single-particle Green's functions in our $g^2$ approximation has the (symbolic) form[5,12]

$$G = \tilde{G} + \tilde{G}MG - \tilde{F}^{(1)}M^{(h)}F^{(2)} - \tilde{G}M^{(1)}F^{(2)} - \tilde{F}^{(1)}M^{(2)}G, \tag{6}$$

$$F^{(2)} = \tilde{F}^{(2)} + \tilde{F}^{(2)}MG + \tilde{G}^{(h)}M^{(h)}F^{(2)} - \tilde{F}^{(2)}M^{(1)}F^{(2)} + \tilde{G}^{(h)}M^{(2)}G,$$

where $\tilde{G}$ and $\tilde{F}^{(2)}$ are the well-known Gor'kov Green's functions, which, in contrast to the standard case,[2,3] contain $\tilde{\varepsilon}$ and $\tilde{\Delta}$ from which the contributions of the corresponding $M^i$ have been removed.

Next, following Ref. 7 we represent the mass operators $M$ and $M^{(h)}$ as a sum of parts which are even and odd as a function of energy, for example, $M = M_{(e)} + M_{(o)}$. Then, determining the excitations of an odd-mass nucleus as the poles of the Green's functions, we obtain from the system (6) the formal expression for these energies[5]

$$E_{\lambda\eta} = \sqrt{\varepsilon_{\lambda\eta}^2 + \Delta_{\lambda\eta}^2}, \tag{7}$$

where

$$\varepsilon_{\lambda\eta} = \frac{\tilde{\varepsilon}_\lambda + M_{(e)\lambda}(E_{\lambda\eta})}{1 + q_{\lambda\eta}} \quad \text{and} \quad \Delta_{\lambda\eta} = \frac{\tilde{\Delta}_\lambda + M_\lambda^{(1,2)}(E_{\lambda\eta})}{1 + q_{\lambda\eta}}, \tag{8}$$

and $q_{\lambda\eta} = -M_{(o)\lambda}(E_{\lambda\eta})/E_{\lambda\eta}$. Here the index $\eta$ is the number of the solution of the system (7)–(8). Here the difference from Ref. 7 lies in the fact that by introducing the unobservable, or refined, quantities $\tilde{\varepsilon}_\lambda$ and $\tilde{\Delta}_\lambda$ we avoid taking the $M^i$ into account twice.

A relation between the phenomenological and refined quantities can be obtained from relations (7) and (8). Since the experimental single-quasi-particle energies which we are using should correspond to the dominant (having the maximum spectroscopic factor) levels, the refinement should be such that after the Dyson equations (6) are solved one solution should correspond to an experimental value and the level should remain dominant. These experimental single-quasi-particle energies serve as input data for our entire problem. Using this condition and relations (8) we obtain

$$\varepsilon_\lambda = \frac{\widetilde{\varepsilon}_\lambda + M_{(e)\lambda}(E_\lambda)}{1 + q_\lambda(E_\lambda)}, \quad \Delta_\lambda \equiv \Delta_\lambda^{(1,2)} = \frac{\widetilde{\Delta}_\lambda + M_\lambda^{(1,2)}(E_\lambda)}{1 + q_\lambda(E_\lambda)}, \tag{9}$$

where $E_\lambda = \sqrt{\varepsilon_\lambda^2 + \Delta_\lambda^2}$. The energies $\widetilde{\varepsilon}_\lambda$ and $\varepsilon_\lambda$ in Eqs. (7)–(9) are measured from the corresponding chemical potential $\widetilde{\mu}$ or $\mu$. Solving these nonlinear equations, we can find the refined $\widetilde{\varepsilon}_\lambda$ and $\widetilde{\Delta}_\lambda$, if the phenomenological $\varepsilon_\lambda$ and $\Delta_\lambda$ are known.

We shall now obtain an equation for $\widetilde{\Delta}_\lambda$. For this, since in the limit of no QPI ($M^i = 0$) $\Sigma^{(1,2)}$ becomes the standard BCS gap, and generalizing the corresponding analysis in the theory of finite Fermi systems[2] [see Eq. (4)], we write the mass operator as

$$\Sigma^{(1,2)} = \bar{F}^\xi F^{(1,2)}, \tag{10}$$

where $F^{(1,2)}$ satisfies the system of the equations (6). Here $\bar{F}^\xi$ is the amplitude, which is irreducible in the $pp$ channel and should differ from $F^{(\xi)}$ in Eq. (4), since the Green's functions in Eqs. (4) and (10) are different. It can also be represented as a sum of two parts, similar to Eq. (5),

$$\bar{F}^\xi = \bar{W} + gDg. \tag{11}$$

The interaction $\bar{W}$ is assumed to be energy-independent. The Green's function $F^{(2)}$ in Eq. (10) must also be taken in the $g^2$ approximation [first iteration in Eq. (6)]:

$$F^{(2)} \approx \widetilde{F}^{(2)} + \widetilde{F}^{(2)} M \widetilde{G} + \widetilde{G}^{(h)} M^{(h)} \widetilde{F}^{(2)} - \widetilde{F}^{(2)} M^{(1)} \widetilde{F}^{(2)} + \widetilde{G}^{(h)} M^{(2)} \widetilde{G}. \tag{12}$$

From relations (10)–(12), dropping terms of order higher than $g^{(2)}$, we obtain

$$\Sigma^{(2)}(\varepsilon) = W\widetilde{F}^{(2)} + gDg\widetilde{F}^{(2)} + W(\widetilde{F}^{(2)} M \widetilde{G} + \widetilde{G}^{(h)} M^{(h)} \widetilde{F}^{(2)}$$
$$- \widetilde{F}^{(2)} M^{(1)} \widetilde{F}^{(2)} + \widetilde{G}^{(h)} M^{(2)} \widetilde{G}) \equiv \bar{\Delta} + M^{(2)}. \tag{13}$$

Comparing Eqs. (1) and (13) we see that $\bar{\Delta} = \widetilde{\Delta}^{(2)}$, i.e., the refined $\widetilde{\Delta}^{(2)}$ introduced above satisfies the nonlinear equation

$$\widetilde{\Delta}^{(2)} = W(\widetilde{F}^{(2)} + W\widetilde{F}^{(2)} M \widetilde{G} + \widetilde{G}^{(h)} M^{(h)} \widetilde{F}^{(2)} - \widetilde{F}^{(2)} M^{(1)} \widetilde{F}^{(2)} + \widetilde{G}^{(h)} M^{(2)} \widetilde{G}). \tag{14}$$

The terms with $M^i$ in Eq. (14) give the desired contribution of the QPI to $\widetilde{\Delta}$, and the term $\widetilde{\Delta}_{BCS} \equiv \bar{W}\widetilde{F}^{(2)}$ describes the BCS pairing mechanism but with an interaction $\bar{W}$ that is different from $F^{(\xi)}$ in Eq. (4).

Therefore two problems must be solved in order to take the QPI into account completely (in the $g^2$ approximation) in the problem of pairing in nuclei: 1) $\widetilde{\Delta}_\lambda$ must be found from the system of equations (9), having determined first the quantities $\varepsilon_\lambda$ and $\Delta_\lambda$ from experiment, and 2) Eq. (14) must be solved for $\tilde{\Delta}_\lambda$ or, more precisely, knowing these quantities, we must find the interaction $\bar{W}$ and thereby determine the contribution of terms with and without phonons to $\tilde{\Delta}_\lambda$.

We have performed the corresponding calculations for the semimagic $^{120}$Sn nucleus. First, using an iterative fitting procedure, the phenomenological $\varepsilon_\lambda$ and $\Delta_\lambda$ were determined, starting from existing experimental data for the neighboring $^{119}$Sn and $^{121}$Sn nuclei (see Ref. 5). The equation (4) was solved using the phenomenological $pp$ interaction obtained in Ref. 6: $F^\xi = -C_0/\ln(c_p/\xi)$, where $\xi$ is the cutoff parameter for summation in the range $\xi - \mu < \varepsilon_\lambda < \xi + \mu$. To solve the system (9) and Eq. (14) we used 21 of the most collective $2^+$, $3^-$, $4^+$, $5^-$, and $6^+$ phonons with energy not exceeding the neutron binding energy, which we calculated for $^{120}$Sn on the basis of the theory of finite Fermi systems[2] (see Ref. 5 for more detailed discussion). On account of computational difficulties in solving the indicated nonlinear equations, the calculations were performed for eight single-particle neutron levels from $1g9/2$ to $3p3/2$ near the Fermi surface. However, this restriction is quite reasonable, since the contribution of the QPI is greatest precisely for such levels.

Determining the parameters of the new interaction $\bar{W}$ in Eq. (14) is a separate and very complicated problem, even if $\bar{W}$ is found from the condition that the quantities $\tilde{\Delta}_\lambda$ obtained from the system (9) are identical to the values obtained by solving Eq. (14). For this reason, here we used the simplest method. The interaction $\bar{W}$ was taken in the same functional form[6] as in Eq. (4), but the parameter $c_p$ was determined from the condition that the average values $\bar{\bar{\Delta}}$ found by solving the system (6) and Eq. (14) are the same. The averaging was done according to the formula

$$\bar{\bar{\Delta}} = \frac{\Sigma_j \tilde{\Delta}_\lambda (2j+1)}{\Sigma_j (2j+1)}. \tag{15}$$

We have obtained the following results. The contribution of $\bar{\bar{\Delta}}_{\text{BCS}}$, which characterizes the BCS mechanism with the new interaction $\bar{W}$, is 74% of the average phenomenological gap, which is 1.42 MeV. Therefore the contribution of the quasiparticle–phonon pairing mechanism is 26%. The contribution from the retarded $pp$ interaction due to phonon exchange (the quantity $(\bar{\Delta} - \bar{\bar{\Delta}})/\bar{\Delta}$) is 31%, and the average contribution to $\bar{\bar{\Delta}}$ from diagrams with the QPI which appear in Eq. (14) is $-5\%$. The latter result is obvious: Just as in the case of the particle–hole channel,[8] the contribution of terms corresponding to diagrams with a ''transverse phonon'' (phonon exchange diagram) and with ''inserts'' (self-energy diagram) are opposite in sign. However, the contribution of diagrams with ''inserts'' is small in our case.

The main result of our calculation is that pairing in semimagic nuclei is of a mixed nature. The BCS mechanism with the refined $pp$ interaction makes the largest contribution to the gap width, while the contribution of the quasiparticle–phonon mechanism, which is mainly of a surface nature, is smaller. If the simple recipes proposed are used to determine a new $pp$ interaction and to estimate the effect by averaging according to Eq. (15), then the QPI contribution will be 26% of the gap observed for $^{120}$Sn. In any case the result obtained must be taken into account in the microscopic description of modern experiments studying low-lying excitations in nonmagic nuclei. For odd-mass nuclei this will be shown in Ref. 5 on the basis of a more phenomenological approach than in the present work, where only the system of equations (9) was solved.

S. K. thanks B. Mottelson and G. M. Éliashberg for a discussion of the results obtained in this work.

*)e-mail: kamerdzhiev@ippe.rssi.ru

Translated by M. E. Alferieff